# Fermi and Lotka: The Long Odds of Survival in a Dangerous Universe[1]

Kent A. Peacock

Department of Philosophy
University of Lethbridge
4401 University Drive
Lethbridge, Alberta, Canada.  T1K 3M4

kent.peacock@uleth.ca

**Abstract**

Fermi's Paradox is the contradiction between the fact that it would seem to be highly probable that there are other technologically advanced species beyond the Earth, and the fact that there is no generally accepted evidence for their existence.  Hanson and Bostrom have proposed that there may be a Great Filter, a survival challenge so lethal that it prevents virtually all species from evolving to an advanced stage.  This paper argues that the Great Filter would be not one single factor, but rather simply the statistics of survival in an always-dangerous universe.  The frequency of species that survive multiple existential threats would likely obey a power law such as Lotka's Law, such that the frequency of survivors would diminish as an inverse power of the number of threats.  Since any species that advances to the point at which it is detectable on an interstellar scale likely must survive a large number of existential threats, by Lotka's Law the number of such survivors would be a very small fraction of the candidate species that evolve on various planets.  Some sobering implications of this picture are outlined.

---

1.     Fermi's Paradox and the Great Filter

This paper will argue that a well-known statistical principle called Lotka's Law could provide some insight into Fermi's Paradox (also called the Fermi Problem). This problem is the contradiction between the fact that there is little or no reliable or generally agreed-upon evidence that humanity has detected or has been visited by extraterrestrials, and the fact that it seems to be virtually certain, on general physical and biological grounds, that intelligent life must exist on many other worlds throughout the universe.

This paper will not exhaustively review the numerous attempts to resolve the Paradox. (Webb [23] presents a useful survey of the possibilities, while Heinlein and Robinson [8] provide a sobering alternative perspective.) Nor will this paper attempt a calculation of the probability of life on other worlds in our galaxy, except to note that recent dramatic results from the Kepler probe and other astronomical observations have only made the Fermi Problem more acute by vastly increasing the probable number of exoplanets in the Galaxy [12]. Instead, the aim here is to understand why highly advanced life is (apparently) rare given that simple life must be rather common throughout the universe.

A paper by Nick Bostrom [2] adds an intriguing twist to the debate over Fermi's Paradox. Following Robin Hanson [7], Bostrom suggests that a reason why life, or at least very advanced life, might be scarce in the universe is that there is a Great Filter that every species encounters at some point in its evolution, some survival challenge so lethally effective and universally pervasive that almost no life can pass through it on its evolutionary path from simplicity to complexity. This survival filter has to be something that would operate with near certainty throughout the universe. Bostrom explains the Great Filter as follows:

> The Great Filter can be thought of as a probability barrier. It consists of one or more *highly improbable* evolutionary transitions or steps whose occurrence is required in order for an Earth-like planet to produce an intelligent civilization of a type that would be visible to us with our current observation technology [2].

The question that worries Bostrom is whether the Great Filter has already operated in the past or will operate in the future. If it tends to operate at some fairly early stage in the evolution of life, then, Bostrom suggests, humanity can breathe a cautious sigh of relief, since we would be on one of the very few planets if not the only planet to have dodged the celestial bullet and we would therefore have a chance of a comfortably long future ahead of us. And we would have reason to believe the Filter has already done its lethal work in the past if we were to find absolutely no evidence of life elsewhere in the universe. If, on the other hand, we find evidence of extinct or primitive life on other nearby worlds such as Mars, but continue to fail to detect evidence of advanced civilizations, that would suggest that the Great Filter tends to act at later stages in the evolution of complex life, and that it therefore likely awaits us in our future. Bostrom (perhaps with tongue in cheek) therefore suggests that we should be glad if we do not find any evidence of life, either extant or extinct, on planets within our own solar system.

One problem with the Great Filter hypothesis is that we presently are not aware of any physical or biophysical basis for any *single* type of lethal effect that could operate with near-100% efficiency at the same stage in the development of life on virtually every habitable planet. Chopra and Lineweaver [3] argue that there may be 'Gaian' bottlenecks in the development of a planetary biosphere: life on many planets may not evolve quickly enough to develop the ability to regulate greenhouse gasses and albedo so as to maintain climate within habitable limits. However, even this proposed survival challenge would probably not act universally enough to serve as Hanson and Bostrom's Great Filter; after all, we know of at least one planet (ours) that so far has (narrowly) evaded the Gaian bottleneck, and it is not plausible that our planet is unique or even all that rare on a large enough scale. Possibly the closest thing presently known that could amount to a Great Filter would be gamma ray bursts (GRBs). Piran and Jimenez [17] argue that there is an appreciable probability that GRBs could harm or even extinguish complex life on Earth (by damaging the ozone layer) and a near-certainty that GRBs would have prevented the evolution of life in the early universe and in the star-dense central regions of galaxies in the present epoch. Again, though, despite the dangers they pose, it is not clear that even GRBs could be as universally lethal as the Great Filter hypothesis demands; furthermore, they could impact a species at any stage in its development.

Of course, we cannot rule out the possibility that there are other survival challenges as yet unknown that could act with the high frequency and lethality necessary to count as a Great Filter. This paper, however, will outline the possibility that if there is anything like a Great Filter, it lies not in some single but presently unknown physical or biophysical principle, but simply in the statistics of survival in a generally dangerous universe. The proposal outlined here would demand revision of some of the factors in the Drake Equation, with the effect of significantly lowering the probability of contact with extraterrestrial intelligence even if primitive life is abundant.

2. The Lotka Curve

In 1926 the mathematical biologist A. J. Lotka published a study of the statistics of scientific publication [9]. Based on two large databases of publications in chemistry and physics, Lotka found that there is an *inverse relation* between the number of papers published and the number of people publishing them; that is, a rather small number of scientists were publishing most of the papers. Lotka showed that the relationship between the number of publications $n$ and the frequency $f(n)$ with which authors have $n$ publications is well-approximated by the power law

$$n^a f(n) = C, \qquad (1)$$

where $a$ and $C$ are constants. As shown by Lotka himself [9] and in Newman [13], $C$ is a normalization constant that appears in any typical power law of this form. The physics is in the exponent $a$; in Lotka's Law it possibly represents, in part at least, complex and hard-to-analyse weighting factors having to do with the innate talent of the individual authors, their varying circumstances, and possible correlations between successive publications. (Successful authors tend to publish more often, partially as a result of their success: 'Those who have, get'.) Taking $a = 2$ and $C = 0.6$ (i.e., 60%), close to the values found by Lotka in his tabulation of scientific publications, we get the familiar Lotka hyperbola showing the relationship between the number of publications and the number of authors with that many publications:

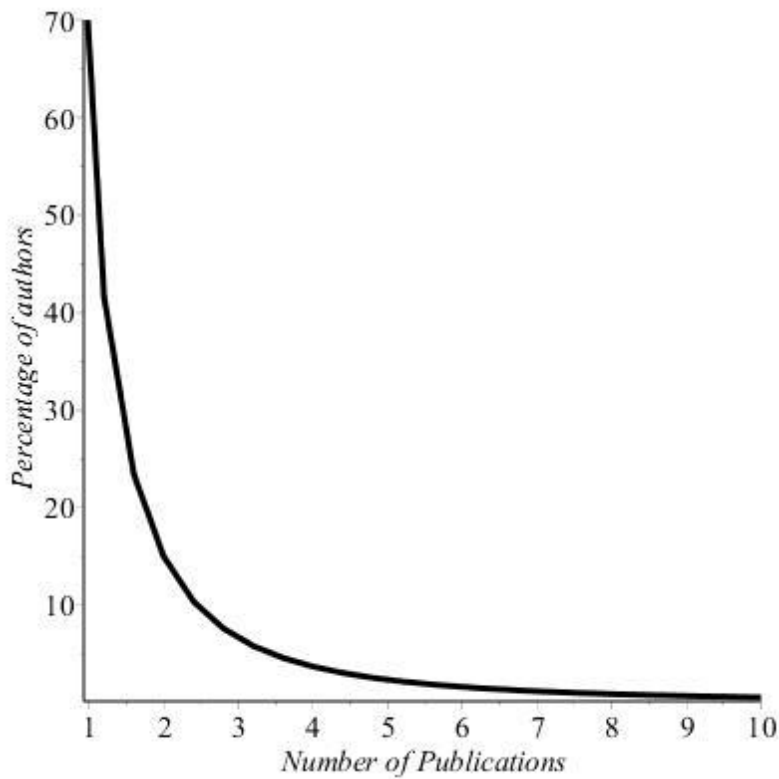

Fig. 1: Lotka's Law of Scientific Publication
(Adapted from Lotka [9])

Charles Murray [11] carried out an extensive study of eminence in many fields of human activity from the arts to the sciences. He showed that if frequency of citations can be taken as a measure of excellence, then accomplishment in many fields also follows a Lotka curve (though with some variation in the values of the exponent).  There also is evidence that combat pilot kills, submarine kills, and tank-combat kills follow Lotka-like power distributions [1].  Ormerod [14] argues that while new corporations may proliferate rapidly in the right conditions, the relatively small number that survive a large number of business cycles is described, again, by a power law.  As Ormerod puts it, *most things fail*.

Lotka's Law is one of several power laws, including Pareto's and Zipf's Laws, that are found to occur widely in nature [4,13,14].  The physical and mathematical basis of power laws is still not fully understood, despite their simplicity and widespread occurrence.  (See Newman [13] for extensive review.)  Murray argues that Lotka-like outcomes will show up most strongly (that is, the curve will peak more sharply to the right) in fields in which high accomplishment is exceptionally difficult [11, pp. 90–106].  However, even though we lack a completely satisfactory general explanation of power laws, all of the cases where Lotka-like curves have been found to apply do have one feature in common:  they involve series of trials with *binary* outcomes.  Lotka's original example of scientific publication is an illustration:  a paper is either published or it is not, with no in-between.  The numerous kinds of citations studied by Murray possess the same rudely binary quality; a book, an artwork, a scientific paper, or a piece of music is either cited or not.  Murray also shows that several kinds of win-or-lose sporting events follow a Lotka curve.  Professional golf is an example:  a golfer either wins a tournament or does not, *tertium non datur*.  Murray shows that only a very small number of outstanding players won most of the PGA (Professional Golfers Association) tournaments held up to 2001.  By contrast, the distribution of quantifiable golfing skills, such as how accurately a player can putt or how far he can drive a ball, tends (according to Murray) to follow something close to a normal (Gaussian) distribution.

There are, to be sure, obvious disanalogies between some of these cases. The ecology of professional golfing is much more forgiving than the ecology of wartime dogfights; we do not execute professional golfers who fail to win a tournament, while fighter pilots who lose a dogfight rarely fly again.  In order to maintain a supply of competitors, the PGA allows runners-up to earn enough money to stay in the game.  But the statistics of tournament *wins* is uncompromisingly Lotkan.  It does not make any difference to the statistics of winning whether the losing competitors get a second chance or have to be replaced after every round.

In sum:  while no attempt will be made in this paper to provide a general analysis of Lotka's Law or similar power laws, we will take it as a working hypothesis that such behavior is frequently manifested in cases where there are a succession of trials with binary outcomes.  Now it will be argued that, given this assumption, Lotka's Law has an obvious application to the Fermi Problem.

3. Lotka Meets Fermi

A species will be subject, from time to time, to *existential threats*: some factor or event that threatens either its biological survival or at least its continuance as a possible 'galactic' species (a species that has the technology and wherewithal to be noticeable by other suitably equipped species on an interstellar scale). The key point is that the survival of such a threat is binary: a species either survives the threat or it does not, and if it does not, it is out of the galactic game, likely permanently in most cases. Thus it is reasonable to investigate the possibility that *the statistics of survival of candidate galactic species would be subject to Lotka's Law or a similar power law*. It is unclear how often existential threats would occur in the career of a species; the Law by itself cannot tell us that. All we can be reasonably certain of is that a species would have to get through several such threats before it can be a player on the galactic scale. Thus, by Lotka there would be only a very small number of such multiple-threat survivors, and thus a very small number of galactic species, even if the initial candidate pool (the number of planets where life evolves) is quite large. It would be as if a PGA player had to win a very large number of tournaments before becoming eligible for a special, once-every-few years tournament.

If this hypothesis is correct, there is no need to invoke a mysterious 'Great Filter'; the universe is rife with existential risks (some of which are reviewed below), and Lokta guarantees that only a few candidate species (relative to the initial size of the 'applicant pool') will survive enough of these win-lose challenges to be noticeable on the galactic scale. Indeed, it is not out of the question that *no* species (no matter how promising) survives to become much more advanced than we are now. Thus, one could agree with Bostrom that the Great Filter is a sort of 'probability barrier,' but it is one that is due not to any one single survival challenge or hazard acting at a particular stage of a species' career but rather the cumulative statistics of on-going survival hazards in an always-dangerous universe. Thus, we should, in fact, be glad if we find evidence of extraterrestrial life, especially relatively complex life, because that means there is a better chance that advanced life can survive the multiple hazards of the universe.

4. Possible Existential Risks, and Possible Responses to Them

It is worth considering a list of possible existential risks that could be encountered by a species roughly like ours, on a planet roughly like ours. This list is by no means exhaustive. We humans would be guilty of potentially fatal hubris if we were to presume that we are presently aware of all of the possible survival threats that our species could face.

Existential risks to a species can be roughly classified into those due to external factors and those that are self-induced. External risks can be further subdivided into *astronomical* risks, *terrestrial/tectonic* risks, and *biological* risks.

Astronomical risks could include instability of a planet's home star, massive solar flares, impacts by asteroids or comets, nearby supernovas, gamma ray bursts (as noted above), or other factors of which we presently have no knowledge. Terrestrial and tectonic risks could include massive volcanism, Gaian bottlenecks, or other sorts of natural climate change perhaps due to lethal nonlinearities in oceanic-atmospheric or ecosystem dynamics. It is worth noting that most of the major and medium-sized mass extinction events throughout Earth's history have been climatological, often (though not always) trigged by greenhouse emissions due to massive volcanism. (A major impact played a role in the terminal Cretaceous extinction 66 mya). Biological risks could include epidemic disease due to emergent pathogens, hostile actions by a technologically superior extraterrestrial species (highly unlikely, one hopes, but not inconceivable), or (again) some factor of which we presently have no knowledge.

The factors mentioned so far could threaten life at any stage in its development on a planet. If and when a species finally achieves a complex technological culture, it becomes subject to a long list of possible self-inflicted existential threats. These are the most immediately interesting because they are the threats that necessarily dominate humanity's attention now, and also the ones that we have the best chance of doing something about (although we do by now have a limited ability to deflect possible impactors given sufficient warning and political will). A list of such threats includes various kinds of ecocide by means of the degradation of supporting ecosystems, leading to factors such as runaway greenhouse warming, loss of biodiversity, loss of topsoil, deforestation, exhaustion of critical resources, or the triggering of emergent disease due to ecological disruption. A recent study by Motesharrei, Rivas and Kalnay [10] shows that under certain conditions economic parasitism by elites can also be sufficient to cause the collapse of a complex society independently of resource exhaustion or environmental degradation (although it would rarely occur in isolation from those factors).

A number of possible existential risks could be classified, at least *prima facie*, as *behavioural*: they include the risk of nuclear war, the tragedy of the commons (in which a game-theoretic 'grid-lock' forestalls effective cooperative action even when that is physically possible and to greater mutual advantage [19]), and conceivably fanatical religion (as suggested by Arthur C. Clarke [15]).

It might be argued that behavioural factors such as lack of foresight in the use of resources are peculiar to humans and thus not relevant in a discussion of the Fermi problem. In fact, ecology suggests that there is a general tendency for organisms to behave in ways that undercut their own survivability (sometimes, though not always, counterbalanced by symbiotic tendencies [16]). Such self-destructive tendencies are unlikely to be peculiar to human beings, even if they manifest in humans in ways that are peculiarly human. Rather, it is the tendency of all organisms to foul their nests or overshoot their resource base under certain circumstances, and any species that survives long enough to be noticeable on a galactic scale must have found some way of transcending or sidestepping these self-undermining tendencies.

The nature of the crises faced and surmounted could also play a role in determining the probability of future wins, in ways that would be very difficult to analyse in detail. As Ronald Wright notes [24], a species may overcome one survival challenge by means of an innovation, only to fall into a 'progress trap' — a further survival challenge caused by the unintended side-effects of that very innovation. Humanity's present dependency on fossil fuels is a good example of a progress trap in Wright's sense.

No attempt is made here to guess the number of existential crises that a species might be expected to face, or how often such crises would be likely to occur. There is no reason to think that they would occur at regular intervals, so some species (such as *H. sapiens*) might be lulled into a false sense of security by a long (say, 10,000 year) period of relatively benign and supportive ecological conditions. Thus, failure to exercise and implement intelligent foresight must be listed among the most significant of the possible self-induced existential risks.

Existential threats to the evolution of advanced life include not only challenges to the survival of a particular advanced species, but also thresholds that must be passed before it is even possible for a biosphere on a planet to harbour complex life. For example, if a planet becomes locked permanently into a phase where its anaerobes are dominant, then complex technological life might never evolve [22]. Several such evolutionary thresholds have to be surmounted (again, a binary process) before there is any possibility of the appearance of a radio-emitting species on a planet.

6.  Lotka and von Neumann

Several authors have observed that von Neumann self-replicating automata could be used as the basis for an automated interstellar probe. As Bostrom, citing Tipler [20], explains [2],

> A von Neumann probe would be an unmanned self-replicating spacecraft, controlled by artificial intelligence, capable of interstellar travel. A probe would land on a planet (or a moon or asteroid), where it would mine raw materials to create multiple replicas of itself, perhaps using advanced forms of nanotechnology. These replicas would then be launched in various directions, thus setting in motion a multiplying colonization wave.

Since the multiplication of the probe would be exponential, it would seem that such a probe could colonize a whole galaxy (or perhaps infect is a better term) in a time that is very short compared to the time required for the biological evolution of intelligent species. If a technological species were to send out a von Neumann probe, the offspring of the probe might survive much longer than the species itself. Even if it is technically more difficult to produce a von Neumann probe than we currently imagine, it would only take one successful effort in order to fill the whole universe with them. The fact that we have so far not detected any extraterrestrial von Neumann probes can be taken as especially compelling evidence, therefore, that there is no other technological species than ourselves within some very great distance.

The problem with the von Neumann probe hypothesis is that the spread of the probes could well be subject to the same Lotkan limitations as the long-term survival of ordinary biological organisms. Although this requires more study, it is probably safe to say (at least to a coarse approximation) that in biology, unfettered proliferation and radiation are exponential, while long-term survival is described by power laws. The same would likely apply to von Neumann probes: even if their proliferation is in principle exponential, the probes will be subject to all sorts of survival challenges and their spread throughout the universe would be limited to Lotkan-type power laws just as with the spread of species of biological origin. This does not guarantee that no von Neumann probe could succeed in spreading itself rather widely throughout a galaxy. However, it could be almost or just as difficult for a von Neumann probe to spread itself widely in space as it would be for a biological species. Therefore, it is by no means a given that we should have seen von Neumann probes by now, even if someone out there is producing them.

7.  Conclusion and Some Cautionary Implications

If there is a Great Filter, then the most natural account of it is that it is nothing more than Lotka's Law or some similar power law that drastically limits the number of species that survive a long succession of survival challenges. What matters the most to survival in the long run is not the type of survival challenge, but the number of them. The number of species that can be expected to survive $n$ survival trials will go roughly as an inverse power of $n$. This implies that the Filter is something that is neither strictly before us nor strictly after us, but rather a factor which operates all the time. The universe will never cease to be a dangerous place in which to live — although it seems likely that a species that has contrived to disperse itself widely throughout space would, all things being equal, have a greater chance of long-term survival.

Lotka's Law, or some similar power law, is therefore one of the factors that very likely must be taken into account when trying to estimate the probability of encountering an advanced species. This could be done via a modification of the Drake Equation, possibly by revising downward the parameters $f_i$ (the fraction of life-bearing planets on which advanced life emerges), $f_c$ (the fraction of planets that develop civilization detectable at interstellar distances), or $L$ (the length of time such civilizations release signals or artifacts) [21]. While it is generally understood that these quantities must be relatively low, if the hypothesis suggested here is correct then they must to be further, and perhaps drastically, reduced by a 'Lotkan' discount.

It is to be emphasized that pointing out the possible role of a Lotka-like law in limiting the probability of long-term survival does not by itself solve the Fermi Problem in an absolutely conclusive way. If the Great Filter is nothing other than the grim statistics of survival, there might indeed only be a very small number of civilizations in our galaxy or even our Local Group that have survived long enough for us to have a chance of detecting them (despite the enormous number of possible life-bearing planets). However, it is still conceivable that at least one such very long-lived species would have had sufficient time to spread either their signals, their artifacts, or themselves throughout a very large volume of space. The probability of detecting another advanced species cannot be reduced to zero; any "solution" to the Fermi problem can only be probabilistic. However, the fact that long-term survival likely exhibits power-law behaviour suggests that this probability may well be much smaller than we have so far appreciated.

Another important implication of the picture outlined here is that any advanced species that we do happen to encounter would likely be very far out on the right-hand tail of the Lotka curve. Any such species would possess formidable survival capabilities and thus would be potentially very dangerous to humanity in ways that would be hard for us to predict, even if it harboured no hostile intent as such toward us. Therefore, should we ever encounter other advanced species, extreme respect and caution are advised.

In summary: If the hypothesis advanced here is correct then the reason that we have not yet encountered advanced alien life is simply that it is far more difficult than we have so far appreciated for a technological civilization to survive a long time — not because of any one particular hazard, but because the odds are so high of a species' luck running out when faced with a succession of hazards. Indeed, it seems likely that planetary life will typically have to run a gauntlet of multiple survival challenges before it can produce species that can even count as technological. This is certainly consistent with the history of life on earth; it has taken about 3.8 billion years for our planet to evolve life capable of sending radio emissions to deep space, while we have had that limited capability for only about 100 years. By Lotka, the number of planets that reach even this modest threshold must be a tiny fraction of the number of life-bearing planets throughout the universe.

This need not be an occasion for despair, however; it will not usually be a matter of pure luck whether or not a species surmounts an unusually large number of survival trials, any more than it is pure luck when a particular author publishes an unusually large number of papers. To revert to Murray's golfing example: a professional golfer who has won multiple events has done so in important part (though never entirely) because he is a good golfer; although luck always plays a role, good golfers generate their own luck. (We may define a piece of luck as a favourable statistical fluctuation.) The fact that humanity has so far dodged more than one existential challenge could indicate that we might have the qualities it takes to beat the odds for quite some time to come. But it should be a wake-up call, a caution against arrogance, complacency, and over-confidence. (The fact that humanity has so far dodged nuclear annihilation may indeed be a matter of dumb luck [5,18].) In particular, humanity should take the very immediate existential risks implicit in anthropogenic climate change with utmost seriousness [6]. No doubt many species do survive a few extinction threats more or less by sheer luck. Sooner or later, though, intelligent foresight and creativity must become the dominant factors that permit a technological species to survive repeated threats, and humanity's best long-term bet for survival is to promote whatever social and economic conditions tend to foster those virtues.

Acknowledgements
For helpful discussion and advice, the author is indebted to James Byrne, Thomas Heyd, David McDonald, Andrew Patterson, Evan Peacock, Norm Sleep, John Vokey, Byron Williston, members of the Philosophy through Science Fiction class at the University of Lethbridge (Fall 2017), an anonymous referee, and discussants at the 2015 Meeting of the Canadian Society for the History and Philosophy of Science. He is grateful to the University of Lethbridge for support. None of these individuals or organizations are responsible for any errors or omissions in the present work, which is entirely the responsibility of its author. No competing financial interests exist.

**References**
1. J.J. Bolmarcich, "On the Distribution of Combat Heroes", in *Human Behavior and Performance as Essential Ingredients in Realistic Modeling of Combat – MORIMOC II*, Vol. 2, Military Operations Research Society, Alexandria, VA, pp. 658–691, 1989.
2. N. Bostrom, "Where Are They? Why I hope the search for extraterrestrial life finds nothing", *MIT Technology Review*, (May/June), pp. 72–77, 2008.
3. A. Chopra, and C.H. Lineweaver, "The Case for a Gaian Bottleneck: The Biology of Habitability", *Astrobiology*, **16**(1), pp. 7–22, 2016. https://doi.org/10.1089/ast.2015.1387
4. B. Drossel, "Biological evolution and statistical physics", *Advances in Physics*, **50**(2), pp. 209–295. https://doi.org/10.1080/00018730110041365

5. D. Ellsberg, *The Doomsday Machine: Confessions of a Nuclear War Planner*, Bloomsbury, New York, 2017.
6. J. Hansen, *Storms of my Grandchildren: The Truth About the Coming Climate Catastrophe and Our Last Chance to Save Humanity,* Bloomsbury Press, New York, 2009.
7. R. Hanson, "The Great Filter", 1998. Retrieved June 1, 2018, from http://mason.gmu.edu/~rhanson/greatfilter.html
8. R.A. Heinlein and S. Robinson, *Variable Star*, Tom Doherty, New York, 2006.
9. A. J. Lotka, "The frequency distribution of scientific productivity", *Journal of the Washington Academy of Sciences*, **16**(12), pp. 317–323, 1926.
10. S. Motesharrei, J. Rivas, and E. Kalnay, "Human and nature dynamics (HANDY): Modeling inequality and use of resources in the collapse or sustainability of societies", *Ecological Economics*, **101**, pp. 90–102, 2014. https://doi.org/10.1016/j.ecolecon.2014.02.014
11. C. Murray, *Human Accomplishment: The Pursuit of Excellence in the Arts and Sciences, 800 B.C. to 1950*, HarperCollins, New York, 2003.
12. NASA Exoplanet Archive, California Institute of Technology. Retrieved June 1, 2018, from https://exoplanetarchive.ipac.caltech.edu/docs/intro.html#ack
13. M.E.J. Newman, "Power laws, Pareto distributions and Zipf's law", *Contemporary Physics*, **46**(5), pp. 323–351, 2005.
14. P. Ormerod, *Why Most Things Fail: Evolution, Extinction, and Economics*, Faber and Faber, London, 2005.
15. P. Parsons, "Arthur C. Clarke: Interview", *BBC Knowledge*, (Sept/Oct), pp. 41–43, 2008.
16. K.A. Peacock, "Symbiosis in Ecology and Evolution", in K. deLaplante, B. Brown and K. A. Peacock (eds.), *Philosophy of Ecology, Handbook of the Philosophy of Science, Vol. 11*, Elsevier, Amsterdam, pp. 219–250, 2011.
17. T. Piran and R. Jimenez, "Possible Role of Gamma Ray Bursts on Life Extinction in the Universe", *Physical Review Letters*, **113**(23), 231102, 2014. https://doi.org/10.1103/PhysRevLett.113.231102
18. E. Schlosser, *Command and Control: Nuclear Weapons, the Damascus Accident, and the Illusion of Safety*, Penguin Press, New York, 2013.
19. P.G. Sekeris, "The tragedy of the commons in a violent world", *The RAND Journal of Economics*, **45**(3), pp. 521–532, 2014. https://doi.org/10.1111/1756-2171.12060
20. F. Tipler, "Extraterrestrial Beings Do Not Exist", *Quarterly Journal of the Royal Astronomical Society*, **21**, pp. 267–281, 1980.
21. D.A. Vakoch and M.F. Dowd (eds.), *The Drake Equation: Estimating the Prevalence of Extraterrestrial Life through the Ages*, Cambridge University Press, Cambridge, 2015.
22. P.D. Ward, *The Medea Hypothesis: Is Life on Earth Ultimately Self-Destructive?* Princeton University Press, Princeton and Oxford, 2009.
23. S. Webb, *If the Universe is Teeming With Aliens ... Where is Everybody?* (Second Edition), Springer, Cham, 2015.
24. R. Wright, *A Short History of Progress*, Anansi Press, Toronto, 2004.